# Parameters of Chelyabinsk and Tunguska Objects and their Explosion Modes

Yury I. Lobanovsky

*IRKUT Corporation*

*68, Leningradsky prospect, Moscow, 125315, Russia*

E-mail: streamphlow@gmail.com

**Abstract**

This paper describes briefly a mathematical model that relates the parameters of celestial bodies motion in the spheres of activity of the Sun and the Earth with mass-energy characteristics of these celestial bodies and their explosion modes during destruction in the Earth atmosphere, that in turn are linked with phenomena observed on the underlying surface. This model was used to calculate the characteristics of the objects which are causes of Chelyabinsk and Tunguska incidents. Thus, the basic data characterizing these two outstanding phenomena were obtained with using a regular physical-mathematical procedure without any speculative hypotheses and/or assumptions.

It turned out that the size of Chelyabinsk object was almost equal to 200 meters, and its mass is close to 2 megatons. The energy of its explosion was 58 megatons of TNT. The minimum size of Tunguska object was equal to 115 m, mass – 0.4 megatons, while the energy of explosion – about 14.5 megatons of TNT. The generality of the origin of these two celestial bodies, which were cometary fragments, was demonstrated. The article also has a criticism of common now, but misconceptions about Chelyabinsk object. Version of this article appeared in Russian language April 12, 2013 (see [1]).

## I. Introduction

As is known, in the morning February 15, 2013 at about 9:20:30 local time, an explosion of some object which was moving along a gently sloping trajectory with a very high speed was at considerable height in the vicinity of Chelyabinsk near the point at approximately 54.85° north latitude and 61.20° east longitude. This point is located approximately 35 kilometers south-south-west of the center of Chelyabinsk, at which was accepted Lenin Square. This object is now called Chelyabinsk (Russian) meteorite or meteoroid [2]. There was early winter morning, and solar time at the epicenter of the explosion was 7:25:20.

On the vast territory stretching between Zlatoust – the city to the west of Chelyabinsk, Troitsk – the city to the south and Miassky – the village in the northeast, were fixed damage to buildings, broken windows and doors [3]. Total 1,613 injured persons were in the incident, most of them – because of the knocked-out windows. Hospitalization was subjected to different data from 40 to 112 people; two victims were placed in intensive care. Such amount of persons affected by falling of the object from space in historic times has not yet been registered [2]. Distance from the epicenter of the explosion up to the extreme points of the destruction zone exceeds 70 km away at least, and Zlatoust and Troitsk are located at the distance at least 90 km. This means that pressure generated by the shock wave from the explosion at a distance of 90 km from the epicenter, was about 5 kPa (kilopascals), what immediately indicates the explosive energy of tens of TNT megatons (for details, this thesis is described in the section IV of this article).

Not far from the epicenter of the explosion is the city Korkino, at the market of which, at latitude 54.89° north and 61.40° east longitude, at the distance from 13.5 to 14.5 km from the epicenter was filmed very important video [4]. Delay (on this video) of the explosion sound from the flare was 89.5 seconds, and with taking into account the temperature distribution of atmosphere [5] and speeding the explosion wave over the sound [6], it was determined that the slant range to the explosion center was 28.9 km. By movement of the shade from the vertically standing mast on the video, it was concluded that the shadow of mast strived for 0.55 of its altitude in the moment of the flare, resulting in a height of the explosion was 25 – 25.5 km.

Fig. 1 shows the projection of the object trajectory and of the dispersion axis of its splinters on the Earth surface in the final stages of the incident. Azimuth of the trajectory in the standard geodetic coordinate system is approximately equal to 75.5º. This means that the object was moving in general from east to west, veering north on 14.5° (practically the same straight line is displayed in Wikipedia as the most likely trajectory [2]). A large black dot is the epicenter of the explosion. In the middle of a translucent square, located near the epicenter, lies point which is equidistant 90 km from the nearest areas of Zlatoust and Troitsk, where were destructions. It should also be mentioned that in Yuzhnouralsk (at a distance of 45 to 50 km from the epicenter) were knocked out factory gate by the shock wave that was the reason of severely injured of man [7]. In the same place soot was moved from chimneys into houses [7], what also indicates a significant increase in pressure at the time of passage of the shock wave.

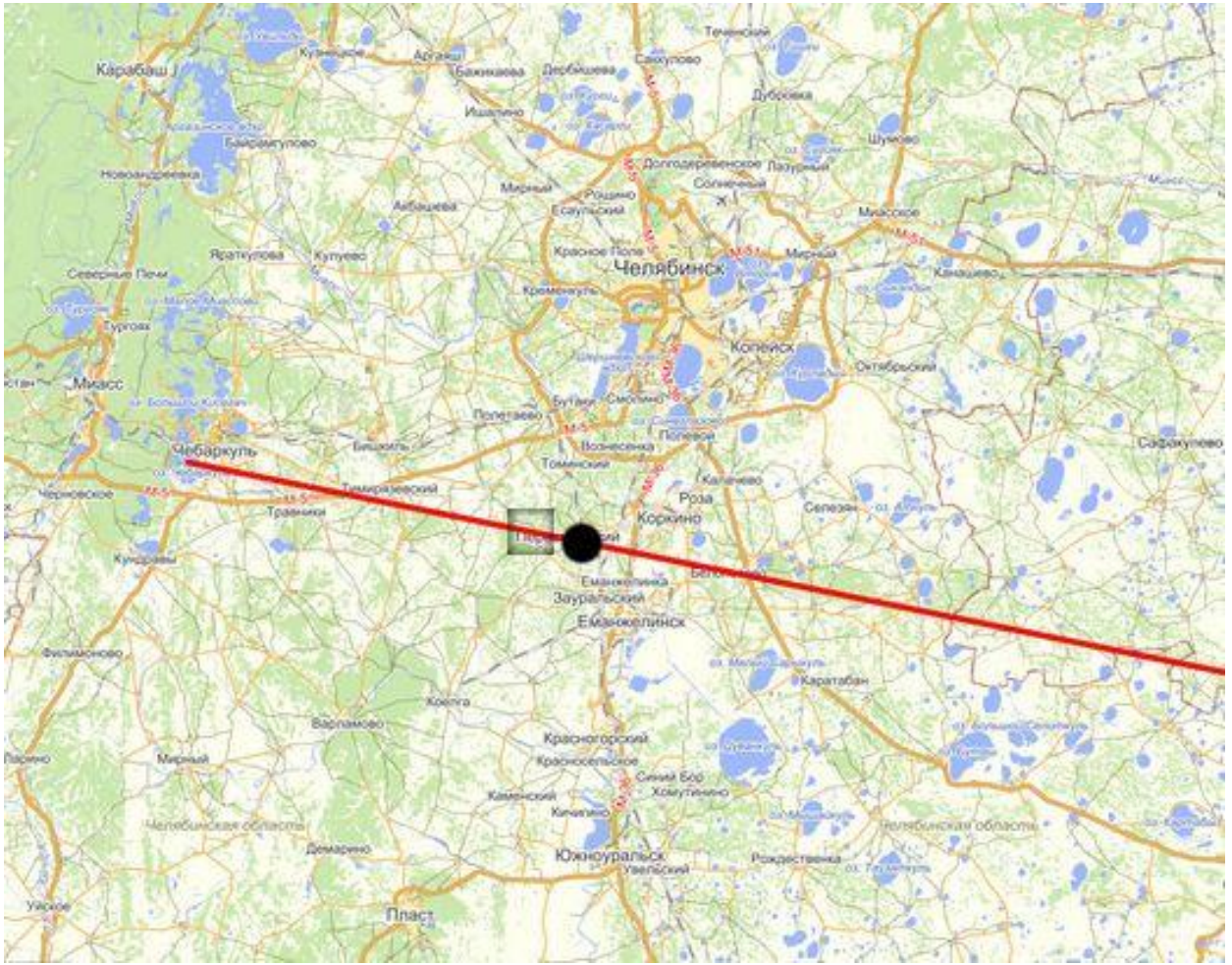

Fig. 1

## II.   Orbits of Chelyabinsk and Tunguska objects

Three independent groups of researchers have identified the orbital parameters of Chelyabinsk object before the collision with the Earth [8 – 10]. These data are represented in three first rows of Table 1 (except for the header line). The linear dimensions (aphelion, perihelion and semimajor axis) are measured in astronomical units, the angle of inclination of the object orbital plane to ecliptic – in degrees. The third group data from the source [10] have considerable scatter, and only mean values of the orbital parameters according to their estimates are shown in the Table 1. Longitude of ascending node (which is uniquely determined by the moment of collision with the Earth object), and argument of periapsis are not discussed here, so they are not shown in the table.

**Table 1**

| Authors and data variants | Aphelion | Perihelion | Semimajor axis | Inclination $i$ (º) | Period of rotation |
|---|---|---|---|---|---|
| **Cooke [8]** | 2.43 | 0.81 | 1.62 | – | – |
| **Lyytinen [9]** | 2.53 | 0.80 | 1.665 | 4.05 | – |
| **Zuluaga, Ferrin [10]** | 2.64 | 0.82 | 1.73 | 3.45 | – |
| **Mean** | 2.533 | 0.810 | 1.672 |  | 2.162 |
| **Based computational** | 2.549 | 0.800 | 1.674 | 3.05 | 2.167 |

As in most similar cases, the orbit calculations are based on random and not very reliable observations and its parameters are determined very approximately. The «internal» differences of the values in the data of the third group with respect to mean values were still greater. Fig. 2 shows the orbit projection on the ecliptic plane illustrating this fact, where the solid line shows the average orbit parameters that are shown in Table 1, and the dotted lines – extreme embodiments from reference [10]. However, despite all this, with some degree of certainty we can conclude that the orbit of the object is almost coplanar with the ecliptic plane. May also be expected that the average values (according to all three groups) shown in the fourth line of Table 1, will not significantly differ from real ones. However, for any specific conclusions it is necessary to raise additional empirical evidence of a general nature.

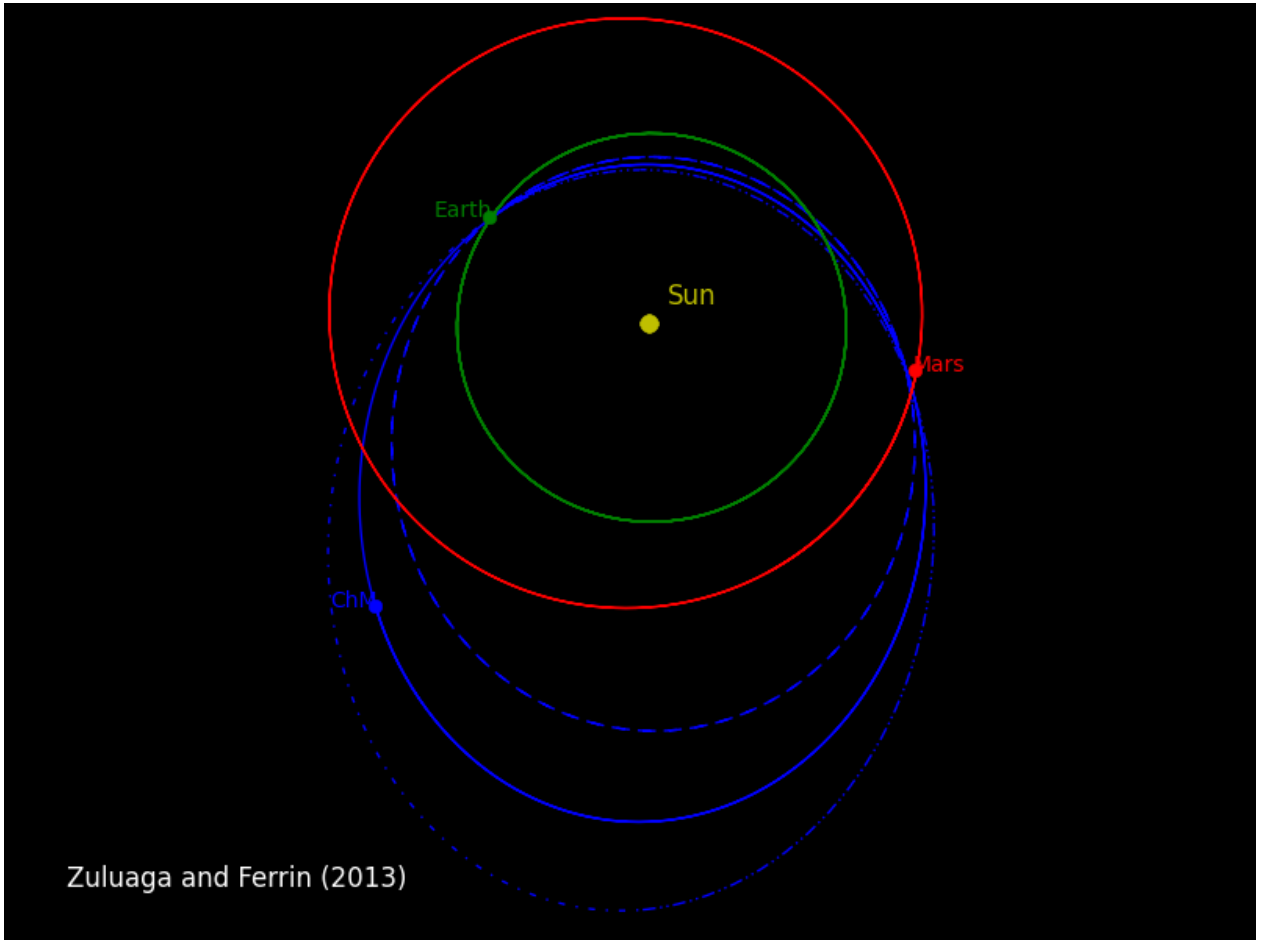

Fig. 2

Now we should pay attention to the fact that the average rotational period of that object is only differ on 0.2 % from the value 2.1667 years. This means that Chelyabinsk object with a high degree of probability was in a 13:6 resonance with the Earth, that is, the rotational periods of the object and the Earth were as whole numbers 13 and 6. Thus, once in 13 years, having made six trips around the Sun, the object again and again was coming closer to Earth until it ran into this planet. In spite of the fact that, as will be shown later, its size on the order exceeds the minimum size of detectable objects in near-Earth space, it has not been fixed by monitoring systems for outer space.

May initially seem that approval about the resonant nature of the orbit of this object was not have been based enough, but there are several asteroids that are in orbital resonance with the Earth, for example, Cruithne [11] and YORP [12], and as they have their own names, they are quite large. We also know that there are spin-orbit resonances between Earth and Moon, Earth and Venus [13], and even Earth and Mercury [14]. More than half a century is known a table in which all the planets in the solar system with an accuracy of 1 % are in orbital resonances with each other [14]. Therefore resonant orbits are not an astronomical exotic, but rather the rule. As is well known, yet Laplace explained the resonant orbits of Jupiter moons with the aid of tidal forces influence. But here, this tidal theory, apparently, cannot explain resonances between orbits of planets and meteorites.

However, this is empirically significant effect, the most obvious manifestation of which is that the meteor showers regularly, once a year or once every few years, invade the Earth atmosphere. And astronomers were found about 1,000 of such flows. However, after the creation of automatic systems to monitor of the space their number is decreased because of the elimination of imaginary meteor showers. And currently have been validated only 64 of them, and further expect acknowledgment about 300 [15]. Those flows of the particles of which are dispersed throughout the orbit, e.g., Perseids intersect the Earth every year. Others, such as Leonids in which particles are concentrated on one part of the orbit, invade into the atmosphere of the Earth once every several years [16]. But, nevertheless, since these meetings are periodically repeated, all these hundreds of flows of particles revolve around the Sun in resonance with the Earth orbit.

Moreover, the Chelyabinsk object collided with the Earth during the second convergence of their orbits, when this object passed perihelion moving away from the Sun (see Fig. 2 at the intersection point of the orbit left of the Sun, signed Earth). Therefore, taking into account the time shift for six months from the date of the first intersection the orbital planes of Chelyabinsk object and the Earth (when, in principle, collision is only possible), the difference between the collision of the objects Chelyabinsk and Tunguska, that have collided with the Earth June 30, 1908, is 104 years, that is exactly 8 periods in 13 years. The difference in a 1.5 month between 30 June and the middle of August – the date of first crossing of Chelyabinsk object of the ecliptic, is easily explained by way of a small difference of the orbits, in the inclinations especially. Consequently, there is every reason to believe that the orbits of these two bodies were close. Using this provision, we can get more information about both celestial bodies and how they ended their existence. In addition, it is quite obvious that if there is more than one of these objects, they must be much, but this issue is discussed in another paper on the subject by author (see [17]). It should be noted that such thoughts begin to come to Western astronomers (D. Kring: «... in fact, it means that somewhere out there are many more «Chelyabinsk» meteorites», see [18]).

**III. Computational model**

After preliminary analysis of the available information, described briefly in the first two sections of this paper, a mathematical model was developed that relates the parameters of motion of objects in the spheres of activity of the Sun and the Earth, with their mass and power characteristics. The latter, in turn, using a Web-based computer program for calculating the consequences of a meteoroid impact on Earth [19] were aligned with the observed phenomena during the Chelyabinsk incident.

Algorithms for calculation the consequences of a meteoroid impact on the Earth have been described in some detail in reference [20]. A model for calculation the parameters of motion of celestial bodies is rather traditional. For a given orbit of the object and the known orbit of the Earth, which, due to its very small eccentricity is assumed for simplicity a circular, the parameters of the object in an elliptical orbit around the Sun are determined at any point from the laws of conservation of energy and impulse-momentum. Then, from the geometrical considerations are calculated angles and speeds in the Sun coordinate system. Further, when the object approaches the Earth, there is a transition to the calculation of its movement within the sphere of activity of the Earth. In this case, the sphere having a zero dimension on the scale of the solar system is infinite in near-Earth space, and solutions in different coordinate systems are sewn through the geometric relationships and mechanical recalculations of speed and energy. The principles which are based in such an asymptotic approach to the description of bodies' motion in the central gravitational fields are described, for example, in [21]. A similar approach has previously been used by the author in [22].

Proceeding to consider the movement of the object in the gravitational field of the Earth there is the problem of computing the so-called impact parameter – the length of the perpendicular drawn from the object velocity vector to a straight line parallel to it and passing through the center of the Earth (see [21]). This parameter determining the motion of an object relative to the Earth can be calculated through known geographical coordinates of the point in which ends of its flight and angle of the flight path. To do this, we need to provide twice the pivoting of the initial system of geographical coordinates. The first twist is performed to account for the inclination of the axis of rotation of the Earth relative to the plane of the object trajectory. The second twist is performed so that the plane of the object trajectory was in the equatorial plane of the new coordinate system. Then, the problem of spatial movement of the object near the Earth passes into the problem of its flat hyperbolic motion, which is described through just one angular parameter – the azimuthal angle $\varphi$. Since there are 2 branches of the hyperbola, there are 2 sets of angles that are providing a second pivoting of the coordinate system. However, from the condition, that the point of explosion is closer to perigee than the point of entry, we may select the only solution that meets the conditions of the problem.

After this remains only problem of action of the Earth atmosphere on the object motion in the final part of its trajectory. This problem is significant merely for small entry angles of the object with relatively long trajectories, one of which was in the Chelyabinsk incident. The solving of this problem is the least trivial part of the algorithm. Because of this, unlike the rest, it is described in this paper in more detail.

There was chosen the simplest embodiment of the accounting method effects of the atmosphere – parameters of trajectory have been calculated for average speed of flight. What is the «average speed» and how it to calculate – this was a major issue in the making of this module of calculation. Based on the data about the explosion of the object (that may be obtained after the calculation of the trajectory), we can calculate the ratio of its remaining kinetic energy just before the explosion $E_f$, equal to the explosive energy $E_e$, to the initial energy $E_0$. Then neglecting the loss in mass of the object in the atmospheric motion

$$v_f = v_0 \left(\frac{E_f}{E_0}\right)^{\frac{1}{2}}, \qquad (1)$$

where $v_f$ – velocity of the object before the explosion (final velocity), $v_0$ – its velocity at the inlet to the atmosphere (initial velocity).

At small impact angles and small changes in object speed can easily obtain that the height of the trajectory z above the Earth surface to a first approximation is proportional to the square of the change in the azimuthal angle φ:

$$z \sim \varphi^2 \qquad (2)$$

Density of the air is the only parameter which is highly changing during hypersonic flight because of its exponential dependence from altitude. Therefore from the formula (2) we provide that the braking acceleration in a first approximation is described as follows:

$$a \approx -c_1 \exp\left(-\xi^2\right),$$

$$\xi = \left(\frac{z - z_f}{h}\right)^{\frac{1}{2}} \approx \frac{\varphi - \varphi_f}{\varphi_0 - \varphi_f},$$

where $c_1$ is a function of the constants determining the aerodynamic forces and mass object, h is characteristic height of the atmosphere, where air density is changed of e times, the index 0 corresponds to the parameters entering the atmosphere, the index f – to the finish of the flight at the point of explosion.

Thus, to a first approximation, after integration over the angle φ we receive a reduction in the object speed Δv in the atmosphere:

$$\Delta v(\xi) \approx -\Delta v_f \left[1 - \mathrm{erf}(\xi)\right],$$

$$\Delta v_f = v_0 - v_f,$$

where erf (ξ) is the probability integral or error function. It is known that in most part of the interval $0 \leq \xi \leq 1$, the function erf (ξ) is close to linear f (ξ) = ξ, and if ξ > 1.5, it almost goes to the asymptote of f (ξ) = 1. Only in a relatively small neighborhood of ξ = 1, there is a smooth transition of function erf (ξ) from one a nearly linear mode depending on the argument ξ to another [23]. So it's a good approximation for using the corresponding piece-wise linear function, and at that case changes in the rate of loss of the object speed along the path approximately may be described as follows:

$$\Delta v(\xi) \approx -\Delta v_f (1 - \xi) \text{ при } 0 \leq \xi \leq 1,$$

$$\Delta v(\xi) \approx 0 \text{ при } 1 \leq \xi \leq \xi_0$$

This means that on the part of the trajectory from the upper edge of the atmosphere $\xi_0 \geq \xi \geq 1$ flight speed is constant and equal to the initial speed $v_0$, and at $1 \geq \xi \geq 0$ it is linearly changes from $v_0$ to $v_f$, and its average value is equal to half the sum of the initial and end values. Then, in the interval $0 - \xi_0$ is easy to determine the average speed of the object <v> on the atmospheric trajectory through statistical weighting coefficients α and β and through the start and end speed values:

$$<v> = \alpha v_0 + \beta v_f,$$

where

$$\alpha = 1 - \frac{1}{2\xi_0},$$

$$\beta = \frac{1}{2\xi_0}$$

For example, for Chelyabinsk object at altitude of conditional entry into the atmosphere 100 km and height of the object explosion 25.25 km (the average between the two values which are discussed further) and the characteristic height of the atmosphere h = 7.16 km (see [21]), $\xi_0$ = 3.23, and weights are as follows: α = 0.845, β = 0.155.

Thus, all parameters of the process are defined through a required number of equations. And at the known object orbit and height of the explosion and coordinates of its epicenter we may uniquely identify all basic parameters of both the object and the burst which is generated through its destruction. The speed and the angle of inclination of the object trajectory are determined at any point. We may easily to define as the length of the trajectory from the entry point into the atmosphere to the point of air burst and its height. And for a given speed and the angle of entry (path angle at the point of entry), together with the known height of explosion and a peak overpressure on a shock wave at a given distance from the epicenter, we may clearly define the characteristics of the explosion caused by the destruction of this object.

However, from the description of the module that calculates the speed at atmospheric part of trajectory it is clear that in order to apply all these relationships, it is necessary to know the parameters of the explosion, which are required to calculate the final speed of the object through the formula (1). In this case, for their determination, in turn, need to know the parameters of the trajectory. And besides, is not known beforehand the azimuth angle of the object entry point into the atmosphere, as well as the impact distance or length of the atmospheric portion of the trajectory. At the same time, in early of calculation of any event, is not clear even level of parameters with which to begin the process of decision. That is why the procedure, which, as can be seen from the analysis of module descriptions of speed calculations, should be the procedure of successive approximations. And all used algorithms should be simple and quick, so we could make a lot of embedded computing cycles on several parameters.

Note that the meteoroid impact module [19] meets all these requirements. And only for this reason such module of computation of the trajectory's atmospheric portion at low angles of entry ($\delta \leq 10º – 15º$) is used, as simplest from adequate variants at the first phase of the numerical model development. But because of this simplicity we should to pay for the proximity of solutions, which, as shown by calculations and analysis of the results, leads to some underestimation of the entry angle $\delta$ and, consequently, an overestimation of the object density $\rho$. If we named the decision, which does not include the effect of the atmosphere on a trajectory, as «unperturbed», the module using such method of «perturbations» for Chelyabinsk object corrects the entry angle by about 60 % from value which could be obtained with the aid of more the exact procedure. Thus, in this case, this approach underestimates the angle $\delta$ about on $0.6º$. However, by using of additional information we can corrected this uncertainty of the existing method.

There are as well errors in the results, which give the meteoroid impact module, the more that it is focused for a fairly steep entry trajectory. Therefore, assessment of the adequacy of the results obtained using the method described here, should be received from a comparison of the calculated and observed data.

Thus, the above-described mathematical model allowed to move from full of uncertainty and speculations about incidents with the inputs of celestial bodies into the atmosphere to the regular solution of completely certain physical and mathematical problem. If necessary, on the basis of this model and initial approximation of the solutions obtained with its help, we may create computational modules, more accurately describing any of the elements of these phenomena, and get more accurate results. It should also be noted that with the help of «external» correction nearly exact results for the characteristics of Chelyabinsk object have already been obtained, and some of the most important parameters, such as, for example, the energy of the explosion, in the framework of this model are determined practically exact. In addition, it is worth recalling that at sufficiently high entry angles ($\delta > 20º – 25º$) the correction of the data is not necessary.

## IV. Parameters of Chelyabinsk and Tunguska objects and their explosion modes

Period of revolution of Chelyabinsk object and uniquely associated with it length of semi-major axis is calculated from the resonance with the Earth 13:6 with any desired degree of accuracy. However, to determine velocities of the Earth and the object and angles of their intersection, others orbital parameters are needed to calculate the required input conditions in the Earth atmosphere. At the first approximation its orbit is lying in the plane of the ecliptic. Then for computing needs one more parameter is required, as which perihelion was selected that had the smallest dispersion in all three groups of results [8 – 10]. As the table 1 shows that the average value of the perihelion is 0.81 astronomical units, and so computations were made with perihelia $r_p = 0.78$; 0.80 and 0.82. It was assumed in the case of need to continue this series, however, the analysis of the calculation results of made it possible to consider the value of $r_p = 0.80$ as reflecting the reality for the basic tasks. After that, we calculated the noncoplanar orbit with an inclination angle $i = 3.05º$, which is obtained from the correlation with the length of semi-major axis according to [10]. It should be noted that a characteristic feature of parameters for all orbits of Chelyabinsk object is that its radiant rejected on the direction to the Sun at an angle of no more than $12º – 16°$, that is, it came to the Earth from the region of the sky close to the Sun.

A few days ago, more than 2.5 months after writing the previous paragraph in Russian language, the author found another estimate of the orbit of Chelyabinsk object [24], where $r_p \approx 0.775$. There was received (after taking into account this result) that the average value of perihelion $r_p$ on four data sets is 0.80, as was customary in the numerical calculations as main magnitude. Deviation of the average value of the orbital period of the object from the resonance magnitude in this case is practically much the same, but its sign is reversed (+ 0.3 % instead of – 0.2 %). Thus, the agreement of these data and of the calculated parameters obtained for the «optimal» orbit, which corresponds to the minimum size of Chelyabinsk object, became even better.

The results of calculations for height of explosion H = 25.5 km are shown in Tables 2 and 3. In the first of them: var – variant of the calculation of the Chelyabinsk object, $r_p$ – value of the perihelion in astronomical units, i – inclination of the orbital plane of the object to ecliptic plane in degrees, v – entry speed into the atmosphere, taking into account the rotation of the Earth, in kilometers per second, δ – entry angle in degrees, ρ – density of the object in kilograms per cubic meter, D – diameter of the object in meters, m – its mass in megatons, $E_0$ – kinetic energy of the object entering the atmosphere in megatons of TNT, $E_e$ – explosion energy of the object in the same units.

**Table 2**

| var | $r_p$ | i (º) | v (km/s) | δ (º) | ρ (kg/m³) | D (m) | m (Mt) | $E_0$ (Mt) | $E_e$ (Mt) |
|---|---|---|---|---|---|---|---|---|---|
| ChO-1 | 0.78 | 0 | 17.99 | 7.79 | 790 | 170 | 2.03 | 78.3 | 57.7 |
| ChO-2 | 0.80 | 0 | 17.51 | 7.87 | 730 | 176 | 2.09 | 76.6 | 57.6 |
| ChO-3 | 0.80 | 3.05 | 17.52 | 7.98 | 670 | 180 | 2.06 | 75.5 | 57.8 |
| ChO-4 | 0.82 | 0 | 17.02 | 7.63 | 740 | 179 | 2.23 | 77.0 | 57.7 |

As can be seen from Table 2, the increase of perihelion $r_p$ quite naturally leads to a gradual decrease in the speed of the object entry into the Earth atmosphere. The entry angle is maximum, and the density of the object, respectively, is minimum at $r_p \approx 0.80$. The diameter and the mass of the object grow with decreasing speed entry into the atmosphere. The kinetic energy of the input has minimum at $r_p \approx 0.80$, the energy of the explosion remains almost unchanged and its value is approximately equal to 58 megatons of TNT. When height of explosion is 25.0 km trends are similar, mass of the object are the same, but density is higher at 110 – 130 kg/m³, and diameters, respectively, less on 8 – 10 meters. The energy of explosion does not change for different $r_p$ and is of about to 56.5 Mt. Thus, the energy of the explosion in the sky in Chelyabinsk was almost equal to the energy of the most powerful thermonuclear explosion of so-called Tsar bomb (other names – AN602, Kuzka's Mother), performed by the Soviet Union of 30 October, 1961 on the New Earth [25].

However the effect of explosion of Chelyabinsk object on the ground was not so catastrophic due to the high altitude. Table 3 shows the calculated values for these distances of the main factor which is a peak overpressure on the shock wave. Here: var – variant, p – peak overpressure on the shock wave in kilopascals at a distance L from the explosion, measured in kilometers along the ground and demonstrated in the column to the left of the pressure.

**Table 3**

| var | $L_0$ (km) | $p_0$ (kPa) | $L_1$ (km) | $p_1$ (kPa) | $L_2$ (km) | $p_2$ (kPa) | $L_3$ (km) | $p_3$ (kPa) |
|---|---|---|---|---|---|---|---|---|
| ChO-1 | 0 | 14.6 | 20 | 11.5 | 35 | 9.6 | 90 | 5.0 |
| ChO-2 | 0 | 14.7 | 20 | 11.5 | 35 | 9.7 | 90 | 5.0 |
| ChO-3 | 0 | 14.6 | 20 | 11.5 | 35 | 9.6 | 90 | 5.0 |
| ChO-4 | 0 | 14.6 | 20 | 11.5 | 35 | 9.6 | 90 | 5.0 |

Even in the epicenter peak overpressure on the wave could not reach 15 kPa, at the distance of 35 km (roughly in the center of Chelyabinsk), he was already below 10 kPa. Peak overpressure 5 kPa at a distance of 90 km is a boundary condition for the solution of this problem. When a height of explosion is 25.0 km, excess of the peak overpressure in the epicenter would be 0.4 – 0.5 kPa, at 20 km distance – near than 0.3 kPa, and then they would almost compared with the values that are presented in Table 3. The wave with a peak overpressure 5 kilopascals on a flat terrain without shielding by buildings knocks out the windows enough with confidence, at 10 – 15 kPa may be damaged and weak destruction of multi-storey buildings. Still, these issues are discussed in more detail in another article in this series.

We now turn to Tunguska object (TO). It has been suggested in section II that its orbit was very close to the orbit of Chelyabinsk object (ChO). We will deduce from this statement all the possible consequences for the moment, and will look how they are consistent with an array of information on the Tunguska «meteorite» that had accumulated during the century which has passed since its fall. If we will to compare ChO-2 and ChO-3 variants, we can conclude that a small noncoplanarity (angle i ≠ 0) affects weakly the characteristics of the object even with small entry angle. And with a high angle of entry, which, according to reports, Tunguska object had, these differences may lose any significance. Therefore, at first time we will consider the simplest version with i = 0, that coincides with the orbit of ChO-2 variant.

The explosion of Tunguska object occurred June, 30 1908, in the first window of approach with another position of the Earth axis to ecliptic plane and with such velocity vector position of the object, which leads to a mirror image of it relative to the velocity vector of the planet compared to what it was in February 2013. The Tunguska explosion was considerably northerly of Chelyabinsk – its coordinates were: 60.89° north latitude and 101.90° east longitude [26]. Local time of the explosion was 7:14:30, solar time – 7:02:06. All of these factors combine to affect the increase of the entry angle of Tunguska object, which at azimuth of 81° [27] was equal to 51.2° instead of 7.9° for ChO-2 variant having the same orbit. Such steep entry trajectory is considered much easier and faster than that of with small entry angles, and in these calculations don't use the above-described module of calculation the speed at the atmospheric part of the trajectory, since such a very short path practically don't change the speed of the object in the atmosphere before the explosion. Thus, inaccuracy caused by this module, are absent, and the density of the object obtained for steep entry calculations is more accurate than for the flat entry.

As before, for calculations it's necessary to put a boundary condition for the peak overpressure of the shock wave from the explosion. Such an obvious boundary in this case is the line of tree-felling. It's believed that it occurs when the wave overpressure is not less than 30 kPa [19]. For several expeditions that took place over the decades, heroic, without exaggeration, researches of Tunguska incident have set these boundaries. The first complete map of the forest fall was composed in 1962 that is 54 years after the event itself. Due to the specific shape of the spot of tree-felling it was named the «butterfly» [28]. One embodiment of this spot – Fast's butterfly, is shown in Fig. 3. The total area of tree-felling on these data is 2150 km$^2$, which is equivalent to a circle of radius of about 26 km. There is recognized that the «butterfly wings» are the result of the impact ballistic shock wave caused by very rapid flight object at low altitude but not as a result of the explosion.

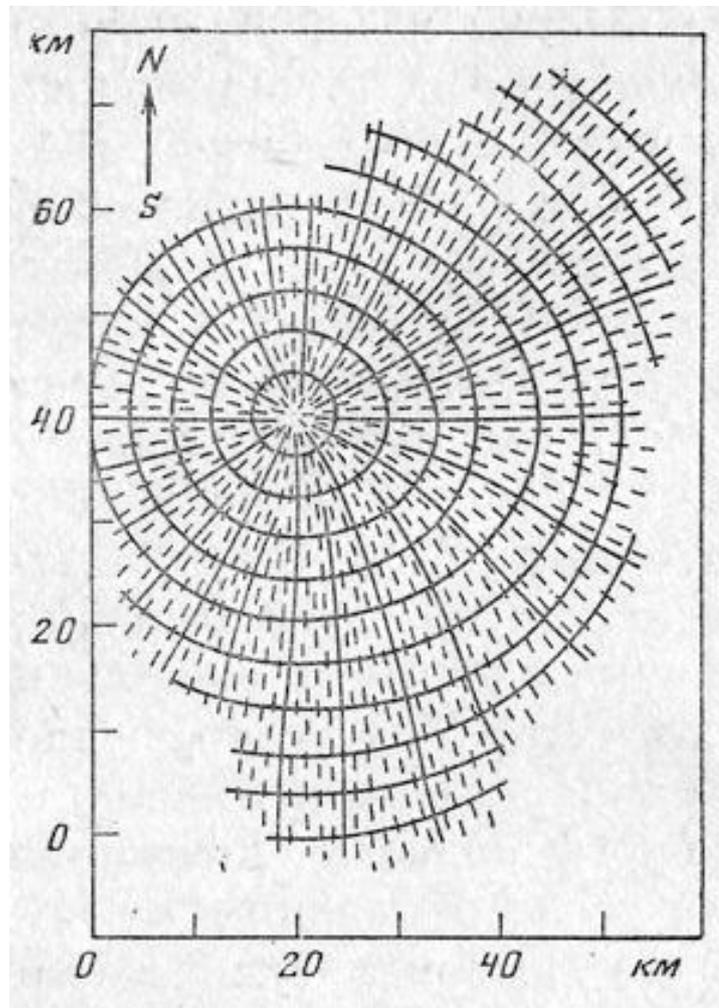

Fig. 3

The explosion takes place as result of the transition of kinetic energy of the object in the heat. It will happen when the movement of small fragments of the object will be practically stopped. Therefore the trace of the explosion effects should not differ too much from the circle. This a priori prediction is confirmed by the results of numerical simulations of Tunguska explosion on a supercomputer lab Sandia [29], see Fig. 4.

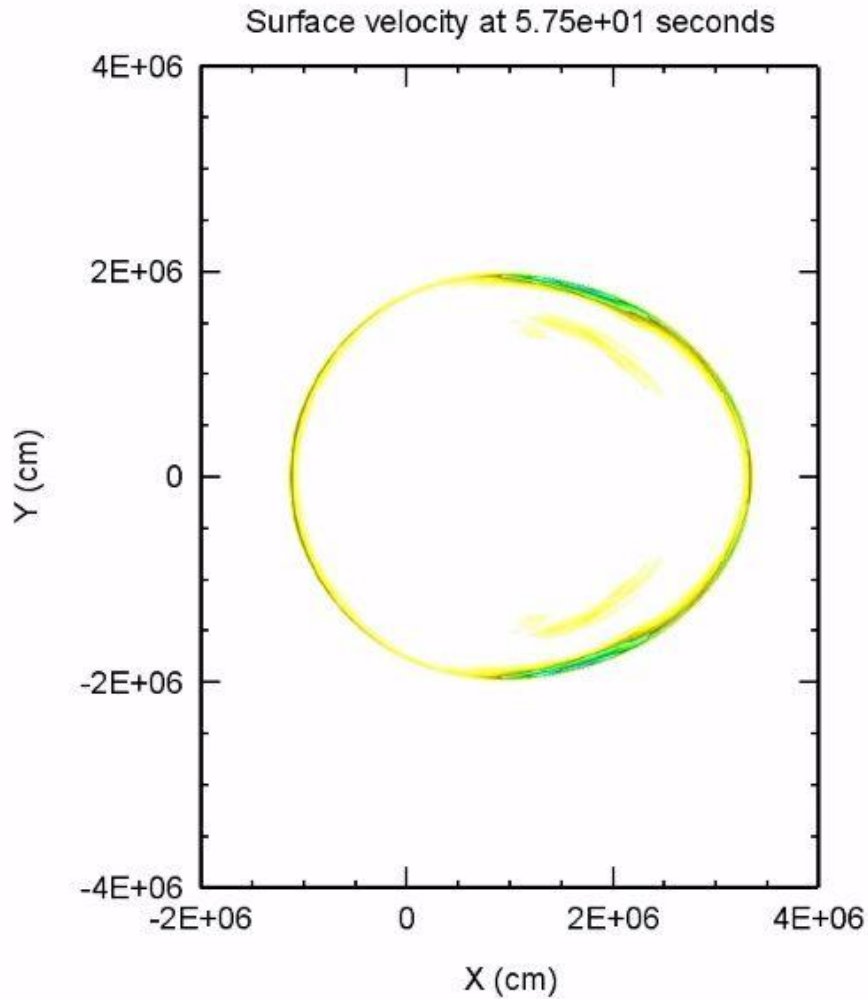

Fig. 4

Comparing Figures 3 and 4, we may conclude that the minimum radius of tree-felling directly from the explosion of Tunguska object was 20.5 ± 0.5 km, and its area is slightly more than 60 % of the total spot area of 2,150 km$^2$, that, according to more recent researches, in the reality was slightly lower.

So, we calculated the TO-1 variant of Tunguska object characteristics with orbit and the density of the corresponding version of ChO-2 Chelyabinsk object. Was also calculated TO-2 variant, differing from the TO-1 in that the density of the object was taken from the ChO-3 variant of characteristics of the Chelyabinsk object with its noncoplanar orbit, which was again shown in the following tables for ease of comparison. The calculation results of all these variants are presented in Tables 4 and 5.

**Table 4**

| var | $r_p$ | i (°) | v (km/s) | δ (°) | ρ (kg/m$^3$) | D (m) | m (Mt) | $E_0$ (Mt) | $E_e$ (Mt) |
|---|---|---|---|---|---|---|---|---|---|
| **TO-1** | 0.80 | 0 | 17.37 | 51.2 | 730 | 97 | 0.352 | 12.7 | 12.5 |
| **TO-2** | 0.80 | 0 | 17.37 | 51.2 | 670 | 101 | 0.361 | 13.0 | 12.9 |
| **TO-3** | 0.80 | 0 | 17.37 | 51.2 | 500 | 115 | 0.398 | 14.3 | 14.3 |
| **ChO-3** | 0.80 | 3.05 | 17.52 | 7.98 | 670 | 180 | 2.06 | 75.5 | 57.8 |
| **ChO-5** | 0.80 | 3.05 | 17.52 | 8.60 | 500 | 195 | 1.94 | 71.1 | 57.8 |

The major differences between the characteristics of Tunguska and Chelyabinsk objects are: almost twice smaller diameter of the first than of the second, in the 5.5 – 6 times less mass and a 4 – 4.5 times less energy of the explosion. However, the peak overpressure of the shock wave at the epicenter of Tunguska explosion is 7 times higher than in the epicenter of Chelyabinsk explosion because of a difference a factor of 3.5 in the heights of these explosions. Next, when the distance from the epicenter increases, a gradual rapprochement between the parameters of two explosions occurs and pressure peaks become equal at a distance of approximately 40 km from the epicenter. At greater distances, much more powerful and more high-rise Chelyabinsk explosion produces more powerful wave.

**Table 5**

| var | H (km) | $L_0$ (km) | $p_0$ (kPa) | $L_1$ (km) | $p_1$ (kPa) | $L_2$ (km) | $p_2$ (kPa) | $L_3$ (km) | $p_3$ (kPa) |
|---|---|---|---|---|---|---|---|---|---|
| **TO-1** | 7.35 | 0 | 98.9 | 19.05 | 30.0 | 35 | 10.5 | 90 | 2.9 |
| **TO-2** | 7.44 | 0 | 98.0 | 19.25 | 30.0 | 35 | 10.7 | 90 | 2.9 |
| **TO-3** | 7.73 | 0 | 97.0 | 20 | 30.0 | 35 | 11.3 | 90 | 3.1 |
| **ChO-3** | 25.5 | 0 | 14.6 | 20 | 11.5 | 35 | 9.6 | 90 | 5.0 |
| **ChO-5** | 25.5 | 0 | 14.6 | 20 | 11.5 | 35 | 9.6 | 90 | 5.0 |

It should be noted that the height of Tunguska explosion was varied, and in Tables 4 and 5 are presented its optimal values, then there are those that correspond to the maximum radius of the tree-felling (see column $L_1$ from table 5). Objects with smaller as well as larger masses tend to reduce this radius. First – due to the lower explosive energy and the latter – due to lower height of their explosions, because of penetration of a larger body through atmosphere to the ground surface. This leads to a sharp increase in the effects of an explosion, but for lesser area.

However, initially supplied boundary condition is not performed for TO-1 and TO-2 variants – 30 kPa overpressure peak on the shock wave is realized at a distance lesser than 20 km from the epicenter. Only variant of TO-3 with a density of 500 kg/m$^3$ produces a result that meets the requirements derived from the empirical description of the consequences of Tunguska explosion. This density is lower by 170 kg/m$^3$ than that which was obtained for ChO-3 variant of Chelyabinsk object. Thus, we may think that there is some lag between the results of calculations of these two, according to the initial assumption, related objects.

However, in section III of this study was indicated that for low angles of entry into the atmosphere, what was for Chelyabinsk object, used in these conditions a simple calculation module of speed leads to some underestimation of the calculated angle of entry and, consequently, to an overestimation of the density of the object. But it is not so for high entry angles because of lack of this calculation method in this case, so there is every reasons to correct the density and the angle of entry into the atmosphere for Chelyabinsk object. This corrected variant is designated as ChO-5 and shown in Tables 5 and 6. It differs from that of ChO-3 with few larger dimensions, but its mass and kinetic energy are decreased by 6 %. And energy of explosion and pressure peaks of the shock wave has not changed at all.

Now consider densities of these objects in a different context. For several decades, it is obvious that the Tunguska object is a fragment of a comet. Accordingly, so is the Chelyabinsk object. Consequently, for the analysis of the adequacy of the solutions, it is appropriate to give a brief overview of the densities of comets. The substance of comets is known to be a composite of dirty snow and ice that have been compacted and many times smelted and frozen. Pollution is, in the main, chondrites, which is the usual stuff of meteorites and asteroids. Since the fraction of chondrites is small, the maximum density of the comet cannot significantly exceed the density of ice which is about 920 – 930 kg/m$^3$ [30, 31]. Various links on the subject [32 – 34] give such varying estimates of densities of comet nuclei – from 100 kg/m$^3$ to 1000 kg/m$^3$, so that it becomes that clear defined and well-founded indications are simply not available. A rough estimate with averaging leads to the value of 550 kg/m$^3$. Data for comet 9P/Tempel 1 with slightly smaller spread (200 – 700 kg/m$^3$) [35], lead to average density of 450 kg/m$^3$.

Some understandings may also be obtained by considering data about snow cover. The density of the old snow on the Earth is 300 – 700 kg/m$^3$ [36] that, on average, leads again to the same level of magnitude about of 500 kg/m$^3$. Of course, the gravity on Earth by many orders of magnitude greater than the levels of gravity on the nuclei of even very large comets, but there in the space the snow is solidified for a period of millions of years but not of several months. So, some analogy between the characteristics of snow on the ground and inside the nuclei of comets may be quite appropriate.

From all these considerations it can be concluded that the level of density of 500 kg/m$^3$ of comet fragment is consistent with the known data on comets and terrestrial snows. Therefore further ChO-5 and TO-3 options with such density are considered as the basic and as correctly reflecting reality.

**V. Discussion of results**

Thus, in the morning February 15, 2013 the fragment of comet has exploded in the sky over Chelyabinsk at a height of 25.5 km. Its size was of approximately 195 m, density – of about 500 kg/m$^3$ and mass – of about 1.95 Mt. Energy of the explosion was 58 megatons of TNT. Over 104.5 years before this, June 30, 1908 the fragment of the same comet has exploded on the Stony Tunguska River, that was much smaller, however, it is still considered the largest celestial body that entered the Earth atmosphere in historic times. Because of unity of origin, it had the same density, but its minimum size was 115 m, and mass – 0.40 Mt. The energy of explosion was about 14.5 Mt, but because of that the height at which this incident has occurred, was 7.7 km, the impact on the underlying surface at that time was much stronger. The calculated data of Tunguska incident are in excellent agreement with those previously obtained by several generations of researchers for decades of work on this problem: the energy of the explosion from 7 to 17 Mt at the altitude of between 6.5 and 10.5 km [37].

It should be noted, however, that the results by Chelyabinsk object are in sharp contrast to those which have been replicated around the world by the media with links to NASA immediately after the incident. The first release from NASA February 15, 2013 reported that Chelyabinsk meteor had size before entering the atmosphere 15 m, mass – 7 kilotons, flight speed was 18 km/s, and energy of explosion was «hundreds kilotons» of TNT [38]. The bases of these estimates have not been specified. Later in the same day a clarification was followed that the size of the object is increased to 17 m, mass – up to 10 kilotons, and the estimate of the explosion energy has grown for some reason already from «30 kilotons » to 500 kilotons of TNT. The arguments for the new estimates are follows: the data «had been collected by five «additional» infrasound stations located around the world – the first recording of the event being in Alaska» [38].

Given that a half of the second degree of the object speed multiplied at the stated mass, and divided the result by 4.18 MJ/kg (specific energy of TNT), any other men than the authors of this release should to receive no more than 390 kilotons of TNT but not 500, it can be concluded that they were in such a hurry that forgot even the law of conservation of energy. In addition, the staff of JPL should to know that energy of final explosion of such small objects is much lower than their initial kinetic energy during the input into the atmosphere as a result of energy dissipation on the trajectory. In this particular case, the calculations lead only to 120 kilotons of explosive energy. With this explosive overpressure peak on the shock wave in Chelyabinsk would be, at least in the more than 300 times lower than observed, and there would be absolutely no damage there.

Obviously, the size of the object could not be determined only with the aid of infrasound stations which record perturbations in the atmosphere. Confusion with the data on energy shows that the size of the object could not have been defined through theirs. This leads only to a single logically valid option – the authors of release have determined the size of Chelyabinsk object as maximum of that they cannot detectable in the near-Earth space with modern automated optical tracking system. It was soon confirmed by the «scientific justification» of this approach [39]. However, none of them have thought that circumsolar angles are not accessible to these systems, but this fragment have flown so – it direction of input was rejected on the direction on the Sun at an angle of about 13.6° (for the ChO-5 variant), see also section IV and/or memorandum [40].

Moreover, soon there were the «additional confirmation» of this erroneous from any point of view of an assessment of explosion energy of Chelyabinsk object – 500 kilotons, see [41]. There have been received these notorious 500 kilotons through correlation between energy of flash light and energy of the explosion (see [42]). However this correlation [42] was made only for one parameter and for energy range of explosions was 0.1 – 1 kilotons. In reality there were big divergences between empirical points and the correlation curve because of influence of many parameters. And authors of work [41] have extrapolated this unreliable dependence as they believed, on 3 orders, and actually even on 5 (!). It is obvious that owing to incorrectness of such extrapolation, it is possible to receive any beforehand specified result.

**Conclusions**

1. The results of calculations by the mathematical model that relates the parameters of celestial bodies motion in the spheres of activity of the Sun and the Earth, with the mass-energy characteristics of these celestial bodies and their explosion modes during the destruction in the atmosphere, turned well matched with the data obtained from observations.
2. Calculations have shown that the size of the Chelyabinsk object was almost equal to 200 meters, and its mass was close to 2 megatons. Energy of explosion was 58 megatons of TNT.
3. The size of the Tunguska object was not less than 115 m, its minimum mass was 0.4 megatons, while the energy of the explosion – about 14.5 megatons of TNT.
4. There was shown the generality of the origin of these two celestial bodies, which turned the cometary fragments.
5. There was demonstrated the fallibility of generally accepted notions about Chelyabinsk incident.